\documentclass[twocolumn]{aastex631}
\submitjournal{ApJS}

\shorttitle{AASTeX v6.3.1 Sample article}
\shortauthors{Zhao C. et al.}
\graphicspath{{./}{figures/}}

\begin{document}
\title{Solar Cycle Prediction Using TCN Deep Learning Model with One-Step Pattern}

\correspondingauthor{Shangbin Yang}
\email{yangshb@bao.ac.cn}

\author[0009-0003-3573-4514]{Cui Zhao}
\affiliation{College of Applied Science and Technology, Beijing Union University, Beijing 102200, PR China}

\author{Kun Liu}
\affiliation{College of Applied Science and Technology, Beijing Union University, Beijing 102200, PR China}

\author{Shangbin Yang}
\affiliation{National Astronomical Observatories, Chinese Academy of Sciences, 20A Datun Road, Chaoyang District, Beijing 100012, PR China}

\author{Jinchao Xia}
\affiliation{Inspur Yunzhou industrial internet Co., Ltd, No.1036, No.1036, inspur Road, High-tech Zone, Jinan City, Shandong Province, China}

\author{Jingxia Chen}
\affiliation{College of Applied Science and Technology, Beijing Union University, Beijing 102200, PR China}

\author{Jie Ren}
\affiliation{College of Applied Science and Technology, Beijing Union University, Beijing 102200, PR China}

\author{Shiyuan Liu}
\affiliation{College of Applied Science and Technology, Beijing Union University, Beijing 102200, PR China}

\author{Fangyuan He}
\affiliation{College of Applied Science and Technology, Beijing Union University, Beijing 102200, PR China}



\begin{abstract}

Human living environment is influenced by intense solar activity. The solar activity exhibits periodicity and regularity. Although many deep-learning models are currently used for solar cycle prediction, most of them are based on a multi-step pattern. In this paper a solar cycle prediction method based on a one-step pattern is proposed with the TCN neural network model, in which a number of historical data are input, and only one value is predicted at a time. Through an autoregressive strategy, this predicted value is added to the input sequence to generate the next output. This process is iterated until the prediction of multiple future data. The experiments were performed on the 13-month smoothed monthly total sunspot number data sourced from WDC-SILSO. The results showed that one-step pattern fits the solar cycles from 20-25 well. The average fitting errors are MAE=1.74, RMSE=2.34. Finally, the intensity of Solar Cycle 25 was predicted with one-step pattern. The peak will occur in 2024 October with a magnitude of 135.3 and end in 2030 November. By comparing the prediction results with other methods, our method are more reasonable and better than the most methods. The codes are available on \href{https://github.com/zhaocui1207/solar-cycle-prediction-by-tcn} {github} and \href{https://zenodo.org/records/14211884}{Zenodo}. 

\end{abstract}



\section{Introduction} \label{sec:intro}

Solar activity is a general term for all activities caused by solar magnetic field, with an average cycle of about 11 years \cite{hathaway2015solar}. At times of intense solar activity, coronal mass ejection (CME), solar wind and flare are enhanced. Such intense solar activities can affect human living environment, as well as the safety of high-tech facilities such as spacecraft, communications and electricity \citep{webb1994solar,bai2003periodicities, dierckxsens2015relationship, lin2023evolutionary}. Therefore, it is very important and meaningful to accurately predict the intensity of the solar cycle. 

Solar activity is essentially caused by the solar dynamo \cite{charbonneau2010dynamo}. And it is a difficult task to predict the intensity of a solar cycle due to the complex internal structure of the solar dynamo. Traditionally, many physical models based on the fundamental theories of physics have been employed \citep{rigozo2011prediction, helal2013early, miao2015prediction, pesnell2018early, upton2018updated}. In the wake of its development for the past few years, deep-learning has been widely applied to solar cycle prediction. \cite{benson2020forecasting} used a combination of WaveNet and LSTM neural networks, select a window size of 528 observations which is 4 cycles × 11 year/cycle × 12 months/year, and a prediction range of 132 observations which is 1 cycle × 11 years/cycle × 12 months/year. Their forecasts show that the upcoming Solar Cycle 25 will have a maximum sunspot number around 106, that the cycle would be slightly weaker than Solar Cycle 24. \cite{lee2020emd} employed EMD and LSTM Hybrid deep-learning model for predicting the sunspot number time series. The prediction was made in 10, 20 and 50 months for the future, respectively. The model predicts that the Solar Cycle 25 peak will occur in 2024 Dec. with the sunspot number around 100. \cite{wang2021solar} used LSTM model to predict the solar cycle. The data from the previous 720 months was input to the model to predict the upcoming 72 months. It can only predict the upcoming 72 months (6 years) at most. Meanwhile they tried inputting data from the previous 10 months to predict the upcoming one month. The model fit better due to the shorter prediction time step, but could not predicted long steps or months. In addition, they made a prediction for the Solar Cycle 25 peak year, and predicted that the peak would occur in 2023 with a magnitude of 114.3. \cite{dai2021sunspot} proposed PSR-TCN model based on phase space reconstruction and temporal convolutional network (TCN) to predict Solar Cycle 25. The study predicted 13-month smoothed monthly sunspot number of the Solar Cycle 25, making the forecast of sunspot number from 2020 January to 2030 December. The maximum sunspot number was 139.55 that would occur in 2024 April. \cite{su2023solar} used N-BEATS model to predict Solar Cycle 25. The input and output window lengths were set to 240 and 120. The sunspot number of Solar Cycle 25 would peak at 2024 February with an amplitude of 133.9 ± 7.2. \cite{su2024solar} proposed the XG-SN integration model. The model predicts solar cycles using the Extreme Gradient Boosting (XGBoost) Ensemble Learning method combined with Sample Convolution and Interaction Network (SCINet) and Neural Basis Expansion Analysis for Interpretable Time Series (N-BEATS). The historical series from the previous 240 time steps were used as input and predicted solar sunspot numbers for the next ten years (120 months). They predicted that the Solar Cycle 25 peak would reach 127.59 around 2024 January. In addition, other deep-learning models were also used for solar cycle prediction \citep{huang2020deep, prasad2022prediction, dang2022comparative, yang2023sunspot}. Models such as CNN+GRU \cite{chung2014empirical}, TCN \cite{bai2018empirical}, SCINet \cite{liu2022scinet}, and iTransformer \cite{liu2023itransformer}, which are based on time series prediction, are suitable for solar cycle prediction as well.  

Although many deep-learning models have been used for solar cycle prediction, most of them are based on a multi-step prediction pattern, very few are based on a one-step prediction pattern. The "step" in pattern refers to the quantity of output values. One-step pattern can predict/output a single value at a time. To predict multiple steps, it is necessary to iteratively apply the one-step pattern. When predicting the subsequent values, the input data will incorporate the previously predicted value. In contrast, the multi-step pattern can predict numerous values once a time. For the multi-step pattern, the longer the data to be predicted, the more likely it is that the accuracy of the subsequent prediction values will diminish. Moreover, multi-step pattern models tend to be more complex than one-step models , and more difficult to train and optimize.  In contrast, one-step models are quicker to train, and their individual predictions tend to be more accurate. More detailed are described in Section 2. In this paper a solar cycle prediction model based on one-step pattern was proposed, with the deep-learning model Temporal Convolutional Network (TCN). A comparison was made between one-step pattern and multi-step pattern in terms of their effectiveness. Finally the intensity for the Solar Cycle 25 peak was predicted based on the one-step pattern. 

The paper is organized as follows: Section 2 introduces the principles of one-step pattern and multi-step pattern. The proposed TCN neural network is given in Section 3. Section 4 describes the data. Section 5 presents the experimental procedure and results, the fitting and prediction results by one-step pattern, the comparison with multi-step prediction, and prediction for the Solar Cycle 25 peak. Finally, the summary is provided in Section 6.

\section{Time Series Prediction} \label{sec:TSP}
Time series prediction arranges data in a chronological sequence and forecasts future data based on the historical data. Solar cycle prediction is a typical time series prediction. It’s categorized into two prediction patterns: multi-step pattern and one-step pattern.

\subsection{Multi-step Prediction Pattern}
In multi-step prediction, $m$ steps historical data are input to the model, and predict n steps future data. As shown in Figure 1(a), each number represents a time step. It indicates that the historical observations of 4 steps ($m$=4) are input to the model, and 3 future time steps ($n$=3) are predicted. The sliding window moves forward one step at a time to adjust the input and prediction steps.

\begin{figure*}
\gridline{\fig{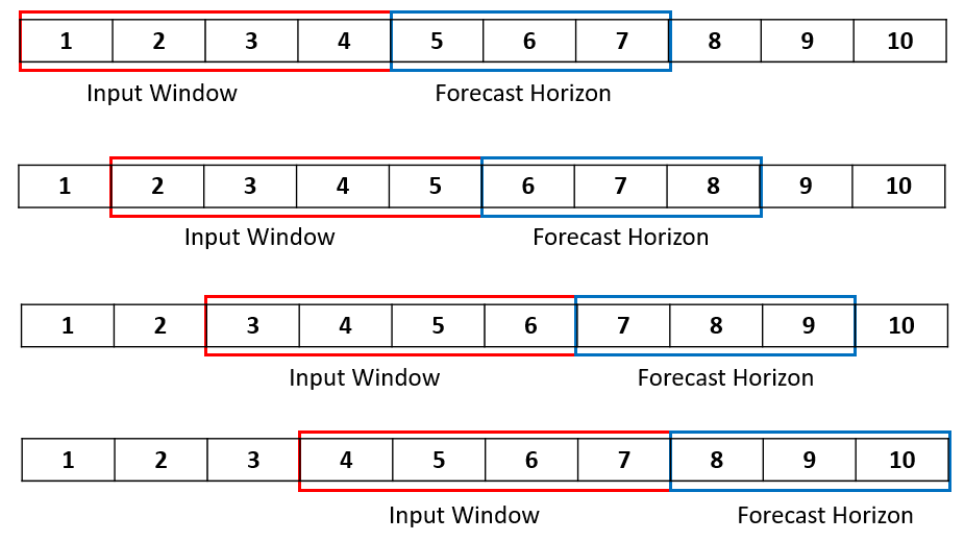}{0.4\textwidth}{(a)}
          \fig{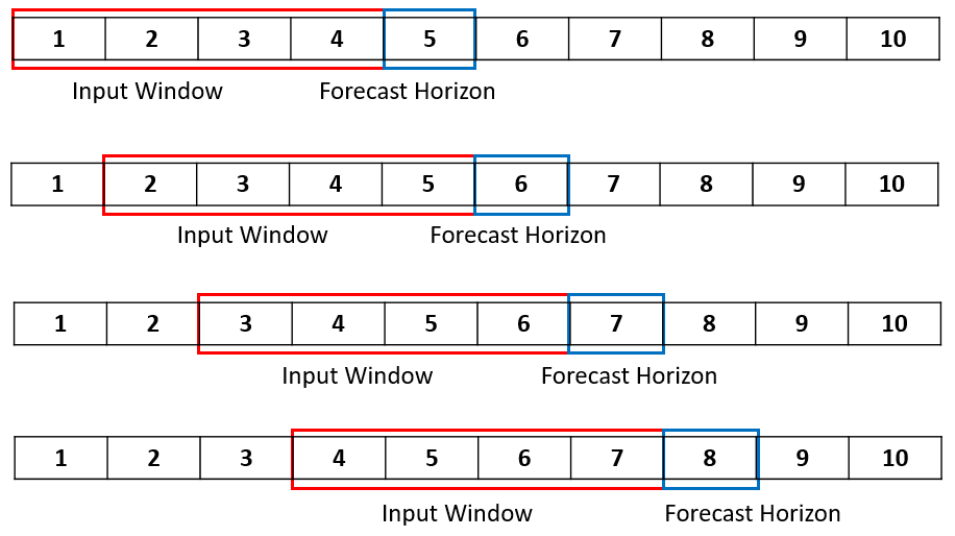}{0.4\textwidth}{(b)}
          }
\caption{Times series prediction: (a) Multi-step prediction. (b) One-step prediction. 
\label{fig:figone}}
\end{figure*}

\subsection{One-step Prediction Pattern}
Figure 1(b) illustrates the principle of one-step prediction. The m steps historical data are input to predict one future value. By autoregressive strategy, one-step prediction is suitable for predicting multiple future data. The principle is that the model generates one output at a time, then adds it to the input sequence to generate the next output. This process is iterated until the entire sequence of n steps is generated. The process is illustrated in Figure 2. The model predicts the output $X_4$ with the input $[X_1,X_2,X_3]$, then $X_4$ is added to the input sequence to make the input $[X_2,X_3,X_4]$, and predict $X_5$. This process is iterated until $n$ future data are predicted, where $n$ is the prediction length. It is noted that the autoregressive strategy is not used in model training process, but only in prediction process. 

In the case of the solar cycle prediction, 1200 months of data are input to predict the subsequent 120 months. The one-step pattern process involves predicting the value for the 1201st month. Subsequently, this predicted value is incorporated into the input data. At this stage, the input data range from the 2nd to the 1201st month, so as to predict the value for the 1202nd month. This iterative process is repeated 120 times in total. In contrast, with multi-step pattern, the values of 1200 months of data are input to predict the next 120 months in a single operation.

\begin{figure*}
\centering
\includegraphics[scale=0.7]{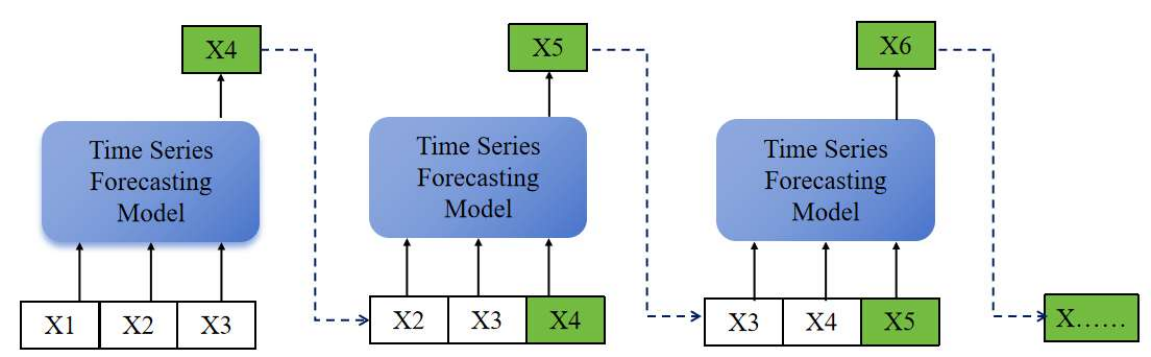}
\caption{Autoregressive strategy. The model generates one output at a time and then adds it to the input sequence to generate the next output}
\label{fig:figtwo}
\end{figure*}

\section{Model} \label{sec:model}
We use the Temporal Convolutional Network (TCN) model as the deep-learning model for solar cycle prediction. TCN is an approach based on the structure of convolutional neural network, which proposed in 2018 \cite{bai2018empirical}. Compared to the classical time series RNN model, TCN model boasts the advantages of higher parallelism, more flexible receptive fields, more stable gradients and smaller memory consumption, and performs well on several time series problems. It consists of two core structures: the causal convolution and dilated convolution. 

\subsection{Causal Convolution}
An output at each moment in the TCN is obtained by convolving only the inputs at that moment and before, so as to ensure the causal constraints in sequence processing. To this end, the causal convolution is provided in the model. As shown in Figure 3, for a value in the upper layer at moment $t$, it depends only on the value in the lower layer at moment $t$ and before. Unlike traditional convolutional neural networks, the causal convolution can’t provide future data, it is a strictly time-constrained model with a unidirectional structure. 

\begin{figure}[ht!]
\plotone{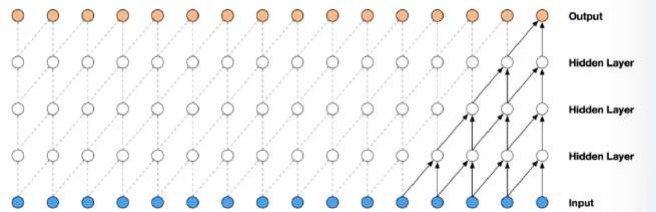}
\caption{Visualization of a stack of causal convolutional layers    
\label{fig:figthree}}
\end{figure}

\subsection{Dilated Convolution}
The prediction length of modeling is limited by the size of convolutional kernels. To capture longer dependencies, many layers would need to be linearly stacked. TCN designs the dilated convolution to increase the receptive field exponentially, without significantly increasing the computational expense. This is shown in Figure 4. Dilated convolution allows for interval sampling of the input at convolution, with the sampling rate controlled by the parameter $d$ in Figure 4. The bottom layer $d$=1 means that every point is sampled for input. The second layer $d$ means that one of every 2 points is sampled for input. The $d$ value increases for higher layers. So the dilated convolution makes the window size grow exponentially with the number of layers. In this way, the convolutional network uses relatively few layers to get a large receptive field. 

\begin{figure}[ht!]
\plotone{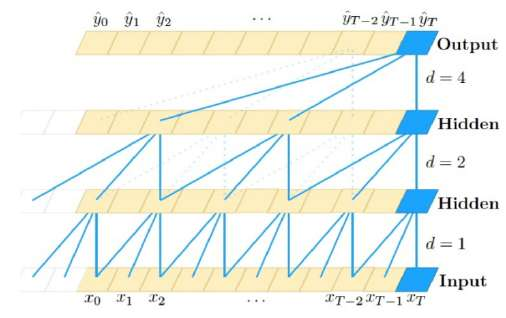}
\caption{A dilated causal convolution with dilation factors $d$ = 1, 2, 4. The receptive field is able to cover all values from the input sequence    
\label{fig:figfour}}
\end{figure}

\subsection{TCN residual block}
Residual connection is an effective and common method for training deep networks. A residual block is constructed in the TCN to replace the convolution of one layer. As shown in Figure 5, a residual block contains two layers of convolution and ReLU nonlinear mapping. WeightNorm and Dropout are added in each layer to regularize the network.

\begin{figure}[ht!]
\plotone{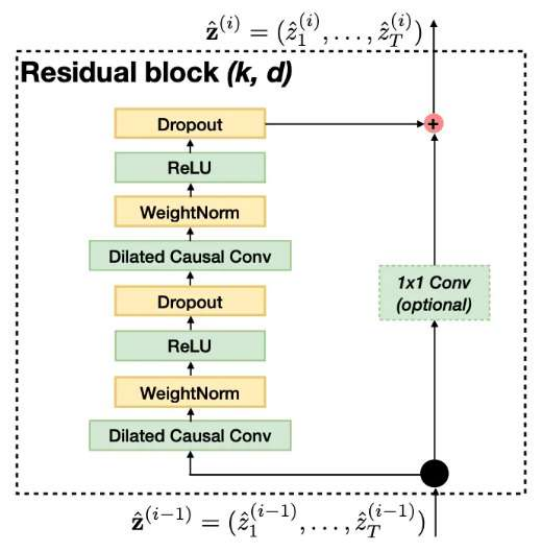}
\caption{TCN residual block    
\label{fig:figfive}}
\end{figure}

TCN features efficient parallel computation capabilities and good processing capacity of long sequential prediction. It reaches or even surpasses RNN model on a wide range of tasks. Its applications include, but are not limited to, processing time series data, such as speech recognition, natural language processing, and other tasks.

\section{Data} \label{sec:data}
The data we used are the 13-month smoothed monthly total sunspot number data from WDC-SILSO, Royal Observatory of Belgium, Brussels. The data timespan is from 1749 January to 2023 September. A total of 3293 usable data are obtained from the data preprocessing by removing outliers less than 0. These data are divided into training set (2470 entries) and test set (823 entries), the training and test set ratio is 3:1. As shown in Figure 6, the black line represents the training set and the blue one represents the test set. The data are finally normalized for model training. 

In solar cycle prediction, the historical sequence $[X_1, ..., X_n]$ is used to predict the future sequence $[X_{(n+1)}, ..., X_{(n+t)}]$, where $n$ is the length of historical data and $t$ is the prediction length. The optimal values of $n$ and $t$ are obtained by experimental parameter tuning. In one-step prediction, $n$=20 and $t$=1, which means the data from the previous 20 months are used to predict the upcoming month. As a comparison, multi-step prediction was also tested, with $n$=1200 and $t$=120, which predicts the upcoming 120 months by the data from the previous 1200 months.

\begin{figure}[ht!]
\plotone{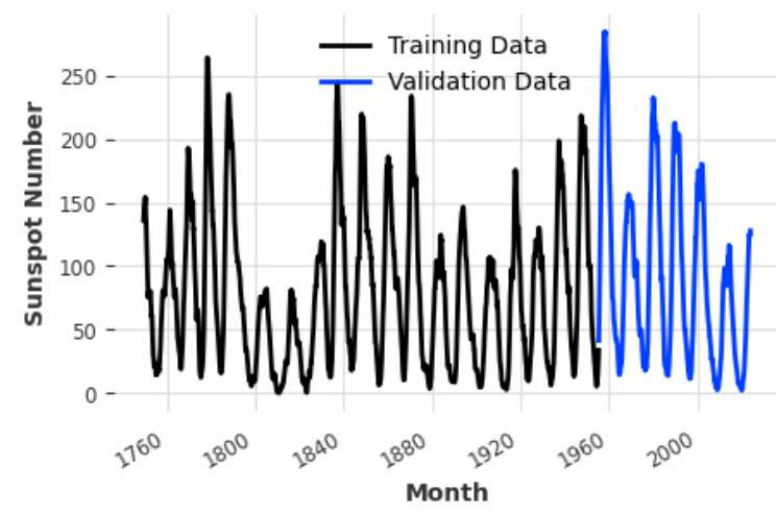}
\caption{Sunspot number data we used, where the black line represents the training set and the blue one represents the test set.    
\label{fig:figsix}}
\end{figure}

\section{Experiments and Results} \label{sec:experiments}

\subsection{Experimental Procedure by One-Step Pattern}
The experiments used TCN model by one-step pattern for training and prediction. The model parameters in experiments were determined by tuning, which are shown in Table 1. All experiments were conducted on a device equipped with Intel(R) Core(TM) i5-7200U CPU. The PYTORCH deep-learning framework was used to build deep-learning models.

\begin{deluxetable*}{cchlDlc}
\tablenum{1}
\tablecaption{Optimal parameters by tuning in experiments \label{tab:tab1}}
\tablewidth{0pt}
\tablehead{
\colhead{Parameter} & \multicolumn2c{Value}  & \colhead{Meaning}  
}
\startdata
input\_chunk\_length & 20 && input length \\
output\_chunk\_length & 1 && output or prediction length \\
n\_epochs & 30 && number of times the model training  \\
dropout & 0 && random loss rate of Dropout in TCN residual block  \\
dilation\_base & 2 && dilation layer by layer with an exponential growth of 2  \\
weight\_norm  & True && normalization layer set true in TCN residual block  \\
kernel\_size & 3 && the size of causal convolution kernel  \\
num\_filters & 6 && number of channels in convolution hidden layer  \\
\enddata
\end{deluxetable*}

\subsection{Fitting Result by One-Step Pattern}
The experiments were first trained on training set. We tuned the input length, and the model fits best when the input length was 20, and the experiments predicted the output at the step 21. The trained model was used to fit the solar cycles from 20-25, respectively. The fitting result is shown in Figure 7, where the black line represents the actual values and the blue one represents the predicted values. Intuitively, the model fits very well.

\begin{figure*}
\centering
\includegraphics[scale=0.9]{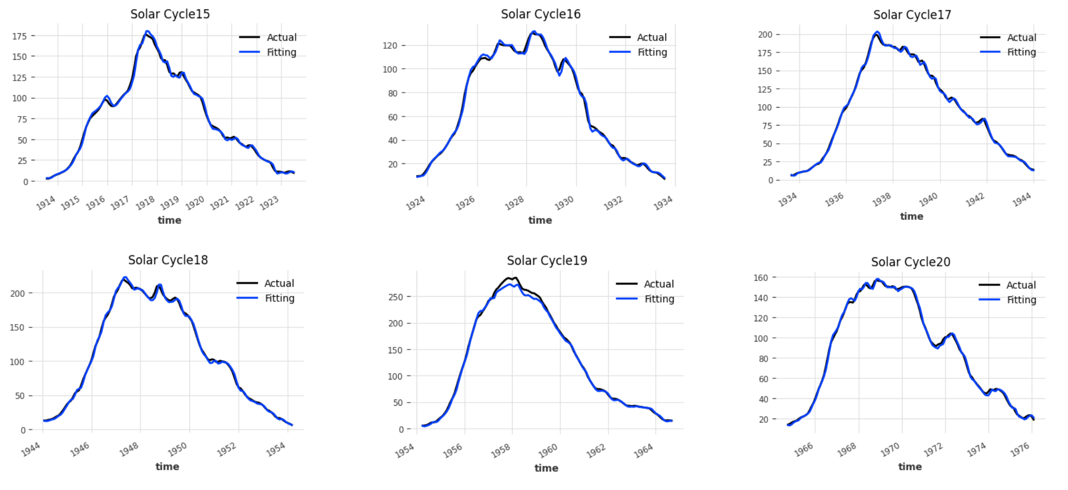}
\includegraphics[scale=0.9]{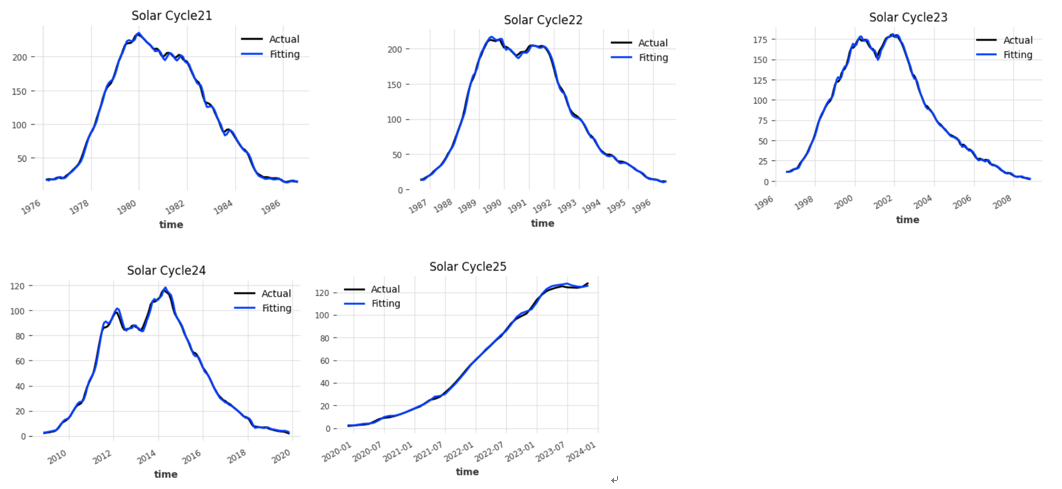}
\caption{Fitting performance of the Solar Cycles from 20-25 by one-step pattern }
\label{fig:figseven}
\end{figure*}

We recorded the differences between the predicted and actual values. The evaluation indicators used here are MAE and RMSE, both of which represent the distance between the expected and actual values. MAE is obtained by calculating and averaging the absolute differences between the predicted and actual values. RMSE is obtained by calculating the square of the differences between the predicted and actual values, and then calculating the square root of the averages. The results are shown in Table 2. Overall, the MAE and RMSE values for the six solar cycles are small, with average MAE=1.74 and average RMSE=2.34. This indicates that the model based on one-step pattern fits very well. 

\begin{table}[t]
\tablenum{2}
\caption{Fitting evaluations of the Solar Cycles from 20-25 by one-step pattern . \label{tab:tab2}}
    \centering
    \begin{tabular}{cccc}
        \hline
        Solar Cycle & MAE && RMSE  \\
        \hline\hline
		 15 & 1.95 && 2.63 \\
		 16 & 1.47 && 1.99 \\
		 17 & 1.82 && 2.33 \\
		 18 & 2.05 && 2.64 \\
		 19 & 3.24 && 4.8 \\
        20 & 1.42 && 1.76 \\
		 21 & 1.88 && 2.48 \\
		 22 & 1.71 && 2.27  \\
		 23 & 1.48 && 1.98  \\
		 24 & 1.09 && 1.60  \\
		 25 & 1.04 && 1.29  \\
		 Average & 1.74 && 2.34  \\
        \hline
    \end{tabular}
\end{table}

\subsection{Prediction Result by One-Step Pattern }
Unlike the training phase, an autoregressive strategy was used in the test and prediction phase. 20 entries of historical data are input, and the next value was predicted iteratively in a loop by the autoregressive strategy until the end of the cycle. Predictions were made for Solar Cycles from 20-25 using the model. Figure 8 demonstrates the predictive effect. The black line represents the actual values and the blue one represents the predicted values. Table 3 records the difference between the predicted and actual values, with average MAE=43.72 and average RMSE=52.92. It can be seen that the predictive effect of the model is not as good as fitting above. This is because the model takes the predicted value as the next input, the input itself carries a certain amount of error, which accumulates as the prediction length increases.

\begin{figure*}
\centering
\includegraphics[scale=0.75]{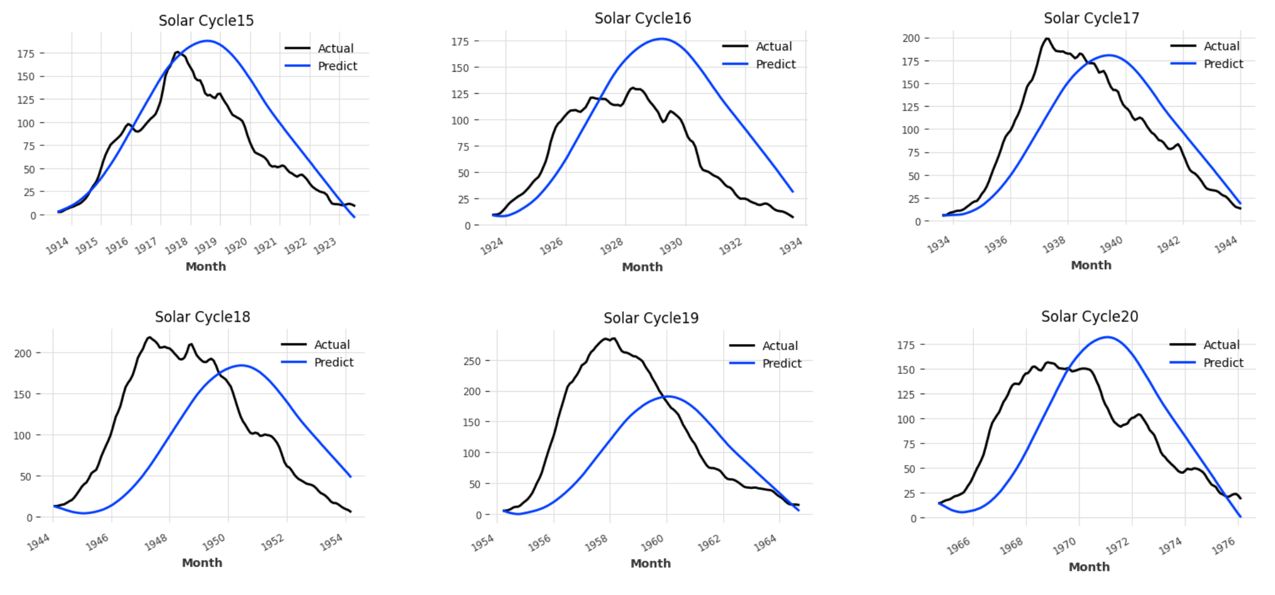}
\includegraphics[scale=0.75]{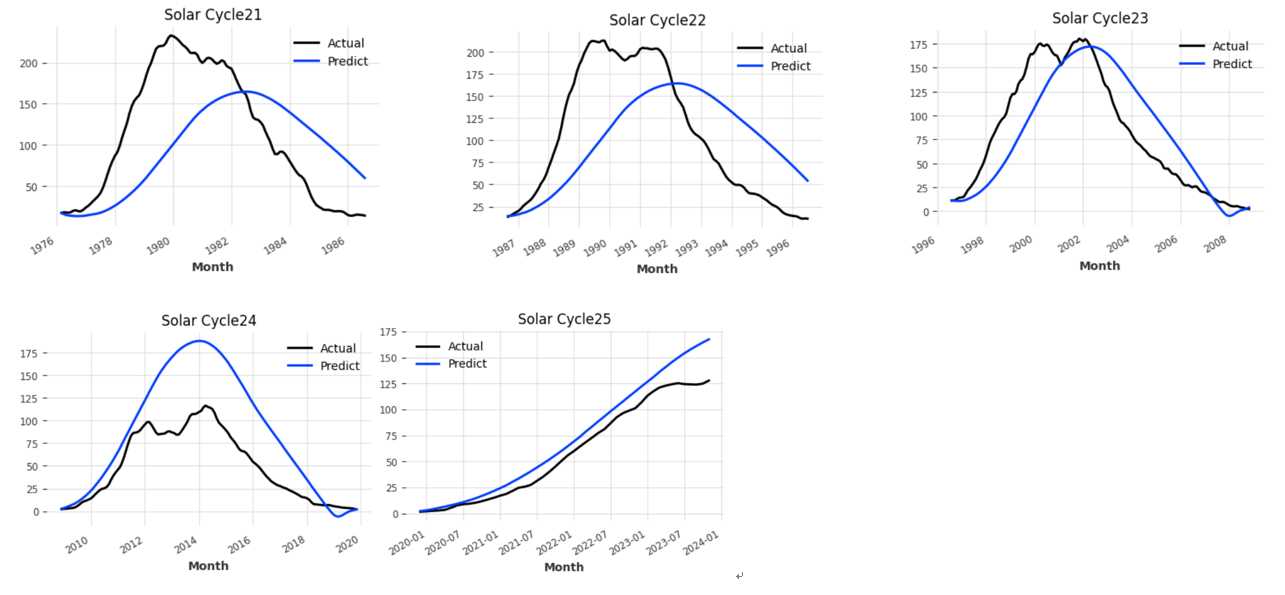}
\caption{Predicting performance of the Solar Cycles from 20 to 25 by one-step pattern  }
\label{fig:figeight}
\end{figure*}

\begin{table}[t]
\tablenum{3}
\caption{Predicting evaluations of the Solar Cycles from 20 to 25 by one-step pattern \label{tab:tab3}}
    \centering
    \begin{tabular}{cccc}
        \hline
        Solar Cycle & MAE && RMSE  \\
        \hline\hline
		15 & 27.08 && 35.17 \\
		16 & 43.9 && 51.11 \\
		17 & 31.98 && 38.43 \\
		18 & 67.76 && 78.53 \\
		19 & 68.43 && 90.3 \\
		20 & 41.62 && 49.95 \\
		21 & 62.16 && 72.85 \\
		22 & 58.23 && 65.86  \\
		23 & 27.99 && 34.59  \\
		24 & 39.54 && 49.8  \\
		25 & 12.27 && 15.57  \\
		Average & 43.72 && 52.92  \\
        \hline
    \end{tabular}
\end{table}

\subsection{Comparison Result with Multi-Step Pattern}
The experiments were also conducted on multi-step pattern in order to compare it with one-step pattern. With 1200 inputs, 120 future outputs were predicted. The input length is the optimal result obtained by parameter tuning, the output length is set to ensure that it can roughly cover a solar cycle. There is no need to use an autoregressive strategy since the training process of multi-step pattern is consistent with the prediction process. For the trained model, the fitting effect is in agreement with the prediction effect.

The experiments fitted Solar Cycle from 20-25 separately and compared the predicted values with the actual values. The results are shown in Figure 9 and Table 4, with MAE=18.52 and RMSE=23.35 for multi-step pattern. It can be seen that multi-step pattern is a worse fitting than one-step fitting, but a little better prediction than one-step prediction.

\begin{figure*}
\centering
\includegraphics[scale=0.8]{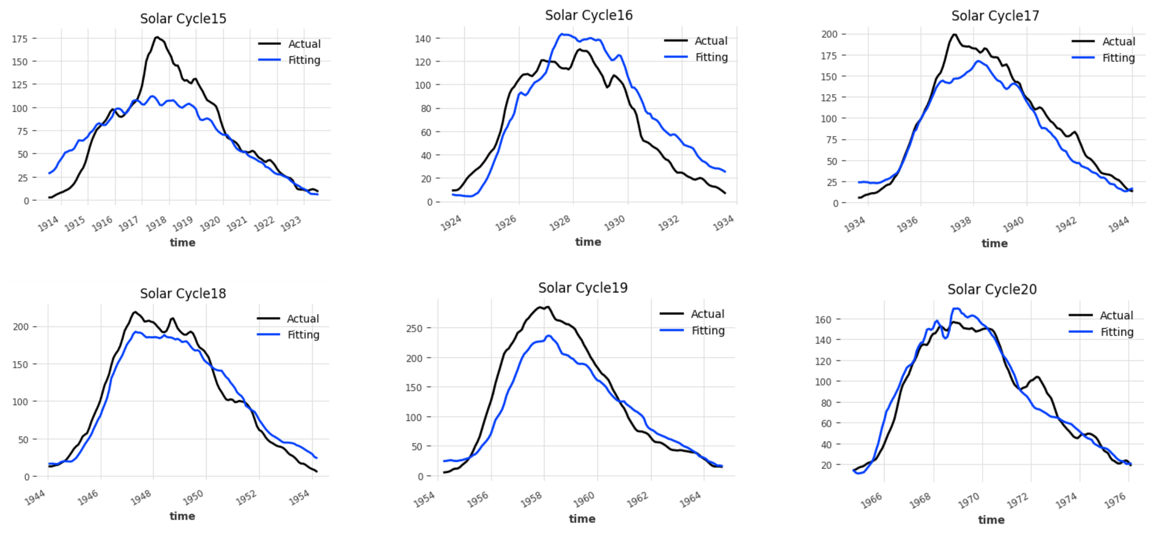}
\includegraphics[scale=0.8]{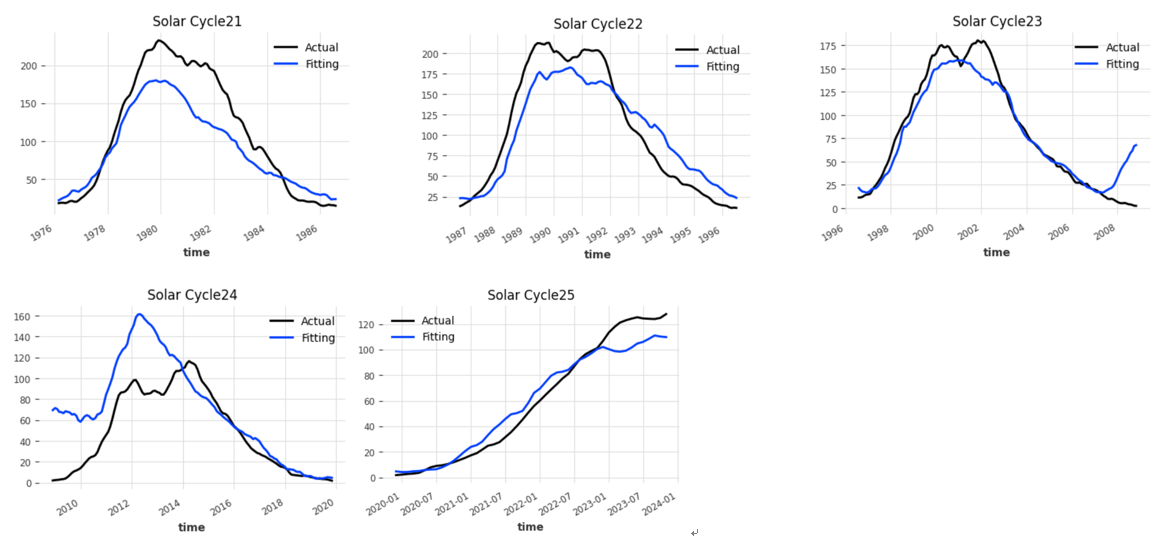}
\caption{Fitting performance of the Solar Cycles from 20 to 25 by the multi-step pattern}
\label{fig:fignine}
\end{figure*}

\begin{table}[t]
\tablenum{4}
\caption{Fitting evaluations of the Solar Cycles from 20 to 25 by multi-step pattern \label{tab:tab4}}
    \centering
    \begin{tabular}{cccc}
        \hline
        Solar Cycle & MAE && RMSE  \\
        \hline\hline
		15 & 18.66 && 26.49 \\
		16 & 18.62 && 19.87 \\
		17 & 14.72 && 18.9 \\
		18 & 14.83 && 16.61 \\
		19 & 28.35 && 36.03 \\

		20 & 8.62 && 11.3 \\
		21 & 28.95 && 35.75 \\
		22 & 25.19 && 28.23  \\
		23 & 12.34 && 18.46  \\
		24 & 25.13 && 34.45  \\
		25 & 8.38 && 10.8  \\
		Average & 18.52 && 23.35  \\
        \hline
    \end{tabular}
\end{table}

\subsection{Prediction of Solar Cycle 25 by One-Step Pattern}
Finally, we predicted Solar Cycle 25 using TCN deep-learning model with one-step pattern. The result is shown in Figure 8. The black line is the actual data as of 2023 September. The blue line is the predicted value by the model starting from 2023 October. Figure 10(a) shows the overall trend for multiple solar cycles, and Figure 10(b) zooms in on the predicted results for Solar Cycle 25. From Figure 10(b), the Solar Cycle 25 peak will occur in 2024 October with a magnitude of 135.3 and will end in 2030 November. This cycle will be slightly more intense than Solar Cycle 24. Meantime, The conclusion has been verified from observations. For example, an  X4.5 solar flare erupted in the active region 13663 on 2024, May 6. Two days later, an X1.0 solar flare erupted in the active region 13665. Another X9.0 solar flare erupted in the active region 13842 on 2024, October 3.

\begin{figure*}[t]
\gridline{\fig{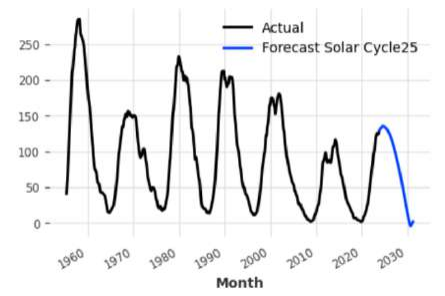}{0.45\textwidth}{(a)}
          \fig{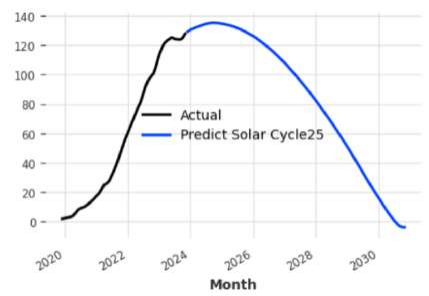}{0.45\textwidth}{(b)}
          }
\caption{Prediction of Solar Cycle 25 by one-step pattern: (a) Forecast trend shown on multiple solar cycles. (b) Enlarge the prediction results of Solar Cycle 25 
\label{fig:figten}}
\end{figure*}

Further, we compared the results of other methods for the prediction of the Solar Cycle 25, as shown in Table 5. The data from SIDC shows that the Solar Cycle 25 peak has actually exceeded 125 by now and is continuing to grow. Therefore, those methods with the prediction results greater than 125 and the peak prediction date after 2023 September are more accurate. It can be seen that the prediction results of our proposed method for Solar Cycle 25 are more reasonable and better than most methods.

\begin{deluxetable*}{cchlDlc}
\tablenum{5}
\tablecaption{Prediction of Solar Cycle 25 by different methods \label{tab:tab5}}
\tablewidth{0pt}
\tablehead{
\colhead{Method from Paper} & \colhead{Deep-learning Model}  & \multicolumn2c{Prediction Pattern} & \multicolumn2c{Peak Value} & \multicolumn2c{Time to Peak} 
}
\startdata
Our Method & TCN && One-step prediction  && 135.3 && 2024.10  \\
\cite{helal2013early} & Statistical method && -------- && 118.2  && 2023 \\
\cite{miao2015prediction}  & Statistical method && -------- && 119.2  && 2024 \\
\cite{bhowmik2018prediction}  & Statistical method && -------- && 118  && 2024(±1) \\
\cite{benson2020forecasting}  & WaveNet+LSTM && Multi-step prediction && 106  && -------- \\
\cite{lee2020emd} & EMD+LSTM && Multi-step prediction && 100   && 2024.12 \\
\cite{mcintosh2020overlapping} & Statistical method && -------- && 233   && -------- \\
\cite{wang2021solar}  & LSTM && Multi-step prediction && 114.3   && 2023 \\
\cite{dai2021sunspot}  & PSR-TCN && Multi-step prediction && 139.55 && 2024.4 \\
\cite{su2023solar} & N-BEATS && Multi-step prediction && 133.9 && 2024.2 \\
\cite{su2024solar}  & XG-SN && Multi-step prediction && 127.59 && 2024.1 \\
\enddata
\end{deluxetable*}

\section{Conclusion} \label{sec:Conclusion}
We provide a new method for accurate prediction of solar cycle and activities. A solar cycle prediction model based on one-step pattern is proposed with a deep-learning TCN model in the paper. By experimenting on the 13-month smoothed monthly total sunspot number data sourced from WDC-SILSO, the following conclusions are obtained: (1) One-step pattern fits the solar cycles from 20-25 well, with an average fitting error of MAE=1.74, RMSE=2.34. (2) Based on the autoregressive strategy, the predictive effect of one-step pattern for solar cycles from 20-25  is gotten, with an average prediction error of MAE=43.72, RMSE=52.92. (3) For comparison, we calculate the fitting effect of multi-step pattern for solar cycles from 20-25, with an average error of MAE=18.52, RMSE=23.35. Multi-step pattern is a worse fitting than one-step fitting, but a little better prediction than one-step prediction. (4) The intensity in Solar Cycle 25 is predicted based on one-step pattern. The Solar Cycle 25 peak will occur in 2024 October with a magnitude of 135.3 and will end in 2030 November. This cycle will be slightly more intense than the Solar Cycle 24. By comparing the actual Solar Cycle 25 observations, it is found that our prediction is better and more reasonable than most methods. Therefore, the solar cycle prediction with one-step pattern proposed in this paper is feasible. Future, we will verify the feasibility of the one-step pattern on more deep-learning models (e.g., LSTM, CNN+GRU, N-BEATs). 

The Jupyter Notebooks to execute the analysis in this paper is hosted at https://github.com/zhaocui1207/solar-cycle-prediction-by-tcn/ and is preserved on Zenodo at https://zenodo.org/records/14211884.

\begin{acknowledgments}

The authors would like to thank the anonymous referee for comments and suggestions that improved the quality of this manuscript. The images and data used in this paper are provided by WDC-SILSO, Royal Observatory of Belgium, Brussels for which we are very grateful. This work is supported by the Academic Research Projects of Beijing Union University (No. ZK20202204, ZK90202106, ZK90202105) ,the Research Topic on Digital Education in Beijing of 2023 (No. BDEC2023619047), the General projects of science and technology plan of Beijing Municipal Education Commission (No. KM202111417002) and Beijing Union University Education and Teaching Research and Reform Project (No. JJ2024Y047).  

\end{acknowledgments}

%





\bibliography{myref}{}
\bibliographystyle{aasjournal}



\end{document}